# Controlled Preparation, Characterization, and Bandgap Modulation of RF Sputtered Antimony Vanadium Oxide (SbVO$_4$) Thin Films.

*Tilak Poudel, Corey Grice, Yanfa Yan, Xunming Deng*

## Abstract

In this paper, RF sputtered antimony vanadium oxide (SbVO$_4$ or "AVO") thin films and its characterization are reported. High purity sputtering targets were fabricated by sintering a mixture of Sb$_2$O$_3$ and V$_2$O$_5$ powders. Thin films were deposited by reactive sputtering at various temperature and argon/oxygen partial pressures. Several growth parameters and surface chemistry were studied by applying numerous optical and electrochemical characterization technique. It is found that SbVO$_4$ exhibits an indirect band gap in the range of 1.89 eV to 2.36 eV and has desirable valence band position to drive water oxidation reaction under illumination. The bandgap depends heavily on stoichiometry of the film and can be modulated by incorporating controlled amount of oxygen gas in plasma environment. Optimized SbVO$_4$ photoanodes was designed and tested for photoelectrochemical water oxidation catalysis. Preliminary studies show that these electrodes possess n-type catalytic behavior in alkaline media. The prepared SbVO$_4$ thin films, with typical film thickness was around 400 nm, contain nanoparticles having the sizes of 10-15 nm.

## Introduction

After the pioneer work by Fujishima and Honda in water photolysis using TiO2 electrode [1] in 1972 [1], extensive efforts have been made to explore oxide and nitrite-based photoanodes TiO$_2$, WO$_3$, Fe2O3, and TaON for the use in photoelectrochemical (PEC) cells [2] [3] [4] [5] [6] [7] [8] because of their relatively high stability in resisting oxidative photo corrosion and their low fabrication cost.

Vanadium antimony based oxide materials have been mostly studied as efficient catalysts for a wide range of selective oxidation and ammoxidation reactions [9] [10] [11] and photoelectrochemical water splitting as well [12] [13]. Prior works suggest that catalytic behavior of these ternary oxides significantly depends on its crystal structure, surface morphology, chemical composition, and its band gap and hence on the preparation conditions [14]. Researches on non-stoichiometric antimony vanadium oxides such as Sb$_{0.92}$V$_{0.92}$O$_4$ [15] [16], Sb$_2$V$_2$O$_9$ [16], Sb$_2$VO$_5$ [17], and modified SbVO$_4$ [18] [19] [20] [21] [22] [23] suggest strong dependence of stoichiometry on the method used to synthesize the film. Even though these oxides are generally more stable in aqueous solutions and inexpensive, but their photochemical activity is often limited by poor carrier separation [24] and their relatively higher band gap. A number of solution-based methods have been explored to make stoichiometric SbVO$_4$ powder [19], thin film deposition with desired stoichiometric SbVO$_4$ photoanodes has remained quite underexplored.

This report describes a novel synthesis procedure for an antimony-vanadium oxide thin film using RF sputtering deposition with high purity homemade targets. Sb and V have different sputtering yields under reactive plasma deposition, so single phase home-made high purity target of SbVO4 in oxygen environment was used to produce the desired phase of SbVO$_4$ by controlling the Sb/V ratio. This approach leads to deposition of uniform SbVO$_4$ thin films with relatively narrow band gap (1.89 – 2.1 eV), desirable for PEC application. We find that a slight oxygen incorporation in gas mixture during sputter deposition lowers the fundamental bandgap by 0.2 eV. Initial photoelectrochemical testing at higher bias potentials indicates an n-type nature of SbVO$_4$ thin film electrode. The surface morphologies of differently prepared thin films were compared and band gap modulation parameters of RF sputtered AVO thin film was also studied.

**Experimental Details**

1) **SbVO4 target synthesis: Preparation of SbVO4 precursor powder**

Precursor powder was synthesized by a high-temperature solid-state reaction method by mixing antimony oxide (Sb$_2$O$_3$, 99.9%) and vanadium oxide (V$_2$O$_5$, 99.9%) keeping the stoichiometric ratios of the metal elements Sb:V nominally 1:1. The mixture was homogenized using a rolling mixer and the powder was transferred to a fused-silica crucible which was placed at room temperature into an ambient-atmosphere electric muffle furnace. A fused-silica plate was placed over the crucible to mitigate the loss of any volatile components (either Sb or V vapor) but without a gas-tight seal to allow excess oxygen to be present during the annealing process. The assembly was brought up to approximately 820 °C over a period of 4 hours and then held at this temperature for another 4 hours. The furnace was then deactivated and allowed to cool down on its own. The resulting powder was strong solid chunk and was dark gray in appearance. This solid material was then crushed and grinded thoroughly using agate mortar and pestle until it became a fine powder. Approximately 20 g of the annealed powder was used to fabricate a target which could be used for RF sputtering depositions. The powder was loaded into a stainless-steel target cup with a 2″ diameter cavity and pressed at room temperature with an applied force of 12 tons for 20 minutes to get the high purity target ready for installation.

2) **Thin Film Deposition: Radio Frequency (RF) Magnetron Sputtering**

The high-purity homemade target was loaded into a custom-built sputtering chamber. Depositions were performed using various substrates (FTO and soda-lime glass) with temperatures ranging from ambient to 270°C, chamber pressures ranging from 5 mTorr to 15 mTorr sustained by supplying a mixture of argon and oxygen gases with the total flow rate of 30 sccm, and RF sputtering powers typically ranging from 40 – 70 W. These conditions resulted in a deposition rate of approximately 1.7-2.5 nm/min. The discussions here are primarily focused on thin films sputter deposited with a 50-Watt RF power at room temperature; and with 6 sccm oxygen and 24 sccm argon, while maintaining a total chamber pressure of 10 mTorr, for qualitative analysis. The substrates were held at room temperature during deposition, and the resulting films were post annealed at 550 °C for 60 mins unless otherwise stated.

3) **Thin film characterization**

The surface morphology and bulk elemental composition (Sb to V atomic ratio) of the thin films were characterized using a Hitachi scanning electron microscope (SEM) with a built-in energy dispersive spectroscopy (EDS) attachment. The ratios of the antimony and vanadium metals in the resulting films were determined using energy dispersive x-ray spectroscopy with Rigaku Cu Kα radiation. EDS measurements for elemental analysis were taken of regions approximately 500 μm x 500 μm in area. Film thicknesses were measured using a DEKTAK profilometer to determine the step height at a tape-masked center and holder frame-masked edges. Bulk crystalline structure of the thin films was characterized using a Rigaku X-ray diffractometer using coupled 2θ Bragg-Brentano mode and a copper X-ray source (Kα Cu =1.54 Å). Phase assignments were made based on the Joint Committee on Power Diffraction Standards (JCPDS) database. The optical absorbance spectrum was measured by a PerkinElmer lambda 1050 UV-vis-NIR spectrophotometer. The transmittance and reflectance of the samples were measured by optical spectrometer using 300 nm – 1500 nm wavelength and the band gaps were calculated using the following relation.

$$\alpha = \frac{1}{t}\ln\left[\frac{(1-R)^2}{T}\right] \tag{1}$$

Where α is the absorption coefficient, t is thickness, and R and T are reflection and transmission respectively. Now, if we plot $(\alpha h\nu)^n$ vs. $h\nu$, the we can get a straight line, the intercept of which gives us the band-gap value.
n= 2 for direct; n=1/2 for indirect transition.

4) **Electrode design and Photoelectrochemical (PEC) measurements**

SbVO4 electrodes were prepared by cutting 4″×4″ thin films deposited on Tec-15 substrates into smaller 0.75″×1.5″ rectangular shapes. The thin films were then partially covered by a non-conducting epoxy resin from the edges to design an electrode, which has a smaller uncovered region in the center for light exposure. The film on one side of these electrodes was then mechanically removed by gentle scratches exposing underlying conducting layer of the substrate and solder with thin indium metal layer for better electrical contact. The PEC characterization was carried out using Voltalab potentiostat, in a three-electrode configuration with an Ag/AgCl reference electrode, platinum mesh counter electrode, and SbVO$_4$ thin film electrode as the working electrode, in a quartz-windowed cell partially filled with 0.5 M KOH solution (pH 13.7) as electrolyte. The recorded potential versus Ag/AgCl ($E_{Ag/AgCl}$) in this work was converted into potential against reversible hydrogen electrode (RHE) using the Nernst equation (2) given below:

$$E_{RHE} = E_{Ag/AgCl} + 0.059 \times pH + 0.1976\ V \tag{2}$$

The system was purged with nitrogen for 30 mins in order to remove possible oxygen dissolved in the electrolyte. Photoelectrochemical response was recorded on both the front-side and back-side illumination through the flat-side of quartz cell. The illumination source was 300 W Xe lamp equipped with AM 1.5 filter. The light intensity of 100 mW/cm$^2$ was adjusted and calibrated using silicon photodiode.

5) **Results & Discussion**
**5.1 XRD and EDS**

This research is primarily focused on obtaining tetragonal SbVO4 by controlling stoichiometric ratio of Sb to V within the range of 1:1 to 1.4:1using reactive sputtering. The amount of oxygen partial pressure during thin film deposition has significant impact on its atomic composition indicating that oxygen can alter the deposition rate of metal atoms. Our results show that under the reactive sputtering, $O_2$ partial pressure affected the sputtering yield of metal atoms. Lower $O_2$ partial pressure increases the sputtering yield of vanadium, higher $O_2$ partial pressure increases the antimony yield. Unlike argon, oxygen forms negatively charged ions in plasma which accelerate towards the substrate and changes the film stoichiometry specially by bombarding with the deposited film and maintaining oxygen content into the bulk. The ratios of antimony and vanadium metals in the resulting thin films were determined using energy dispersive x-ray spectroscopy. The ratio data indicates that film becomes slightly V-rich in complete argon environment and becomes slightly Sb-rich in higher oxygen environment, suggesting that the deposition rate and oxidation state of vanadium depends significantly on presence of oxygen. EDS results with the standard deviation of five measurements obtained for films deposited at room temperature are presented in Figure 1 below.

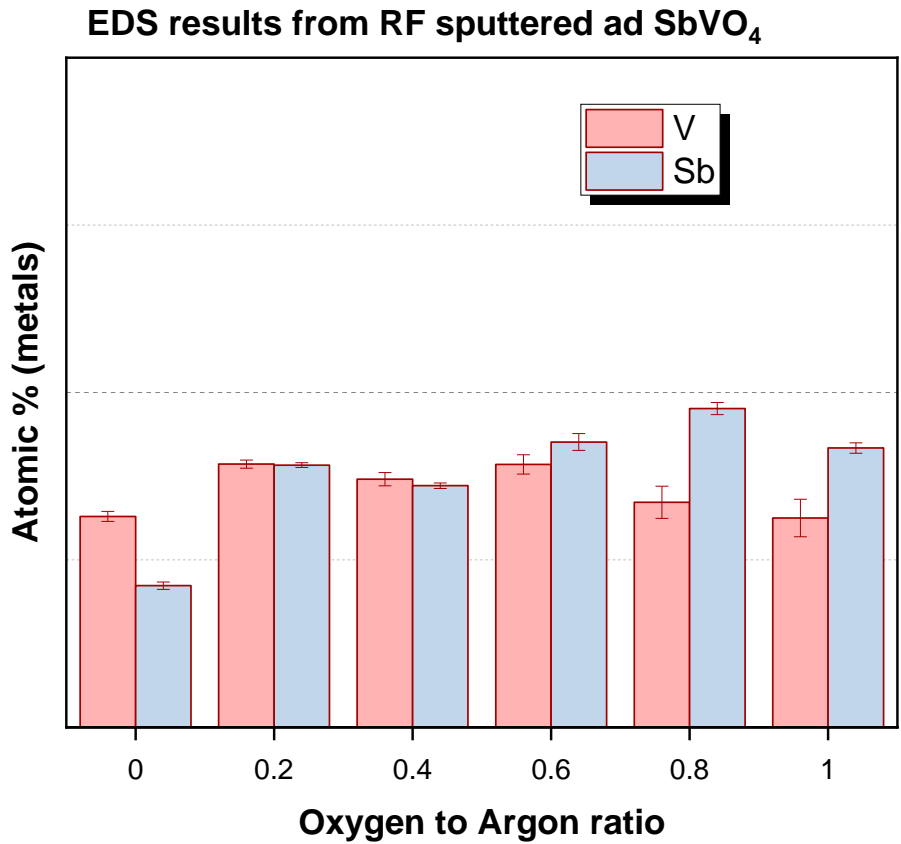

*Figure 1: EDS results from RF sputtered AVO after annealing at 550 °C for 60 minutes*

In the following discussion, we will concentrate on the optoelectronic properties of representative thin films with Sb/V ratio of 1:1 approximately, as obtained by reactive sputtering with 20 % oxygen partial pressure in Ar+$O_2$ environment (a total chamber pressure of 10 mTorr). For each sample, we selected multiple points at different regions within the sample for EDS measurement, and the measured Sb/V ratios at various locations showed that the obtained thin films are compositionally uniform throughout the sample. All samples are approximately 400 nm thick.

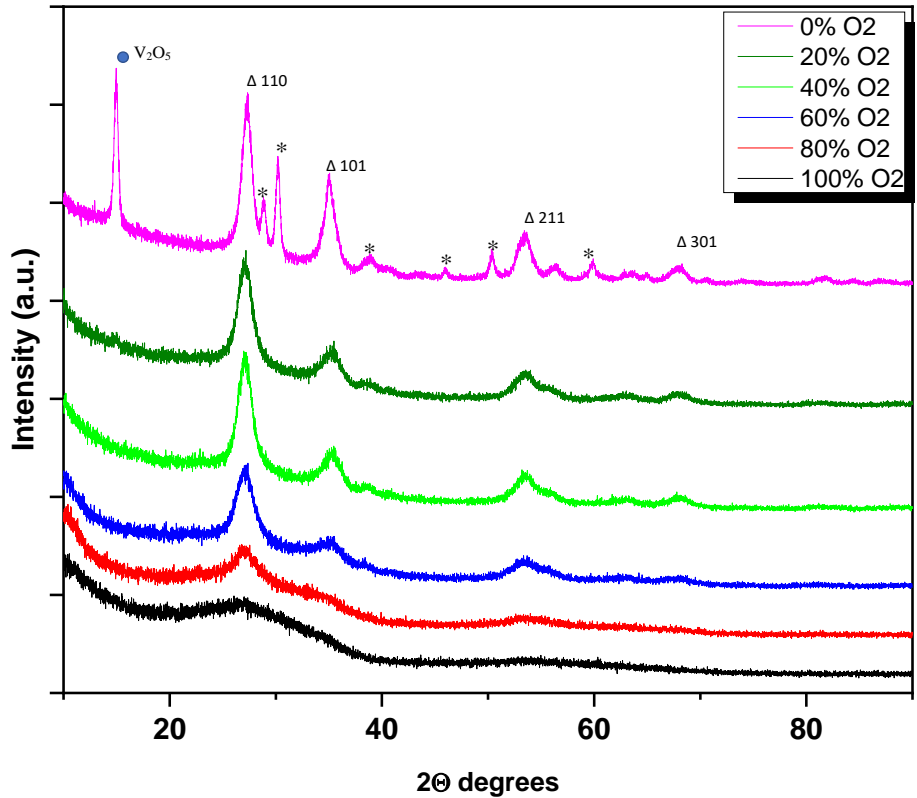

*Figure 2: XRD pattern from various RF sputtered AVO thin films after annealing at 550 °C for 60 min. The sign "*" represents mixed secondary phases.*

Figure 2 shows initial results from x-ray diffraction (XRD) measurements on sputtered deposited AVO films on soda-lime glass substrate. It was determined that the sputtered AVO films were amorphous (see supplementary information) for substrate temperatures up to approximately 270 °C suggesting that a higher temperature is required to transform an initial amorphous film into a crystalline film. Subsequent post-deposition treatments in air at 550 °C for 60 minutes were able to convert a primarily crystal structure into expected tetragonal SbVO4 phase. For the V-rich film, in addition to the pattern of the $SbVO_4$ tetragonal structure, a strong peak appears near 2θ values of 11-12°  this can be assigned to the $V_2O_5$ (002) reflection, showing that excess V can produce phase segregation in the film. We also suspect that the peak at a 2θ value of 28-29° may include combined contributions from identified $V_2O_5$ (003), $Sb_6O_{13}$ and $Sb_2O_4$ (012) reflections. Powder XRD (supplementary information) also indicates that the powder consisted of antimony vanadate, $Sb_2O_4$, which was indexed as a tetragonal structure with lattice constants a = 4.6352 Å and c = 4.6352Å. A secondary trace phase was also present, which was indexed to be an orthorhombic structure with lattice constants a =4.8135 Å, b = 5.4476 Å, and c = 11.7955 Å; the closest ICDD match to this phase was the compound $Sb(SbO_4)$.

XRD measurement indicates that these thin films consisted primarily of antimony vanadate, SbVO$_4$, which was indexed as a tetragonal structure with lattice constants a = 3.164 Å, b = 4.677 Å, and c = 4.677 Å. For the films sputter deposited in a higher oxygen partial pressure, we do not observe evidence of phase segregation in XRD, but reduction in relative peak intensity and shifting of dominant peak slightly towards higher diffraction angle suggest that some growth orientations might be suppressed by additional Sb incorporation. Higher oxygen partial pressure in plasma gas may have reduced kinetic energy of sputtered atoms by scattering during their travel from target to the substrate. Scattering due to presence of oxygen gas along with the substrate bombardment may impact not only the preferential orientation, but also thin film microstructure as well. At room temperature, the deposition rate was determined to be roughly 1.7-2.0 nm/min.

## 5.2 Surface Morphology

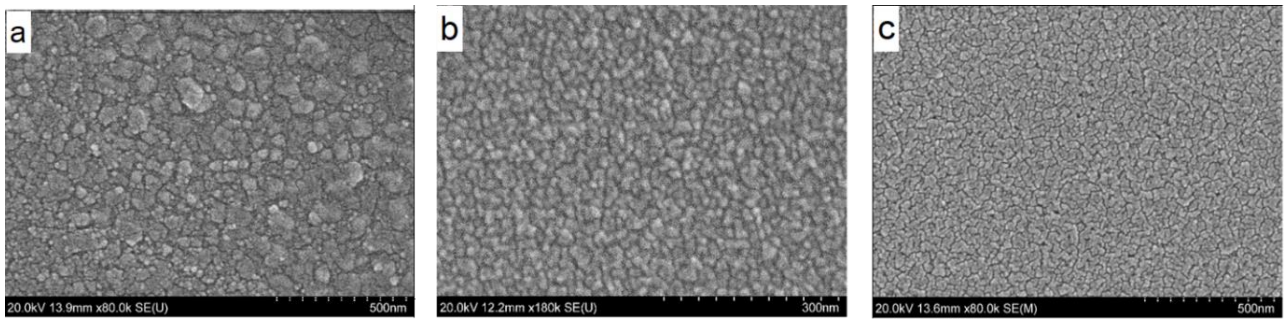

*Figure 3: SEM micrographs of annealed RF sputtered AVO thin films deposited in a) pure argon*

*b) slight oxygen in argon environment c) high oxygen low argon plasma environment.*

Scanning electron microscopy (SEM) was used to determine the morphologies of the sputtered thin films as a function of composition. All films deposited at room temperature were amorphous. After annealing at 550 °C for an hour, all films displayed grains indicating polycrystalline structures as shown in Figure 3 above. The stoichiometric film was characterized by smooth and distinctive grains of a few tens of nanometers in size, whereas slight V-rich film displayed a polycrystalline structure too, but the grains was much bigger but not uniform. XRD also confirmed the presence of the V$_2$O$_5$ impurity phase in those thin films that are deposited in pure argon environment. Sb-rich film (film deposited in high oxygen environment) suffers from significant pinhole issues as in figure c. In high oxygen pressure, sputtered species reach to the substrate with lower kinetic energy and hence reduced mobility may be the leading cause for the formation of pinholes.

Surface morphologies were observed using AFM and are shown in Figure 4 below. The surface roughness of the AVO film was in the range of 15-20 nm.

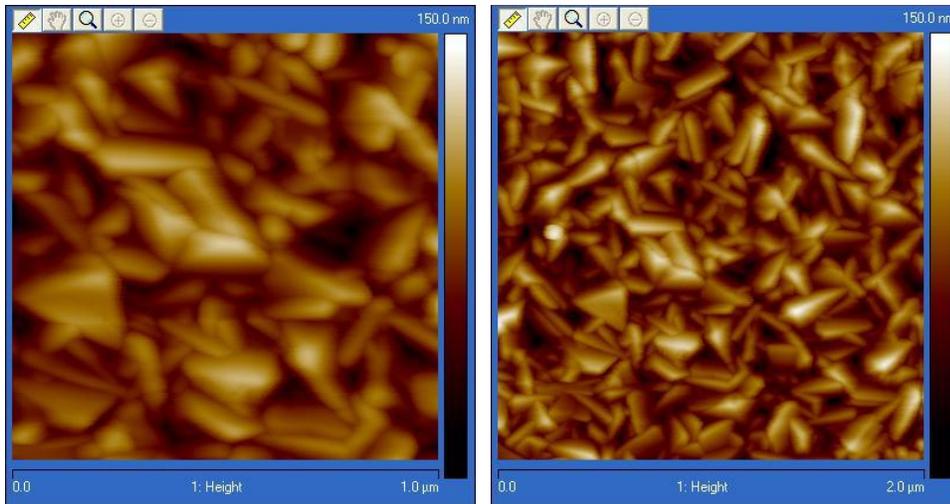

*Figure 4: AFM images of RF sputtered AVO thin films in 20% $O_2$/Ar reactive plasma*

## 5.3 Optical Properties

Transmission and reflection measurements were used in the UV-visible range of the photon spectrum to determine the absorption coefficient and a Tauc plot for estimating the bandgap energy, as shown in Figure 5.

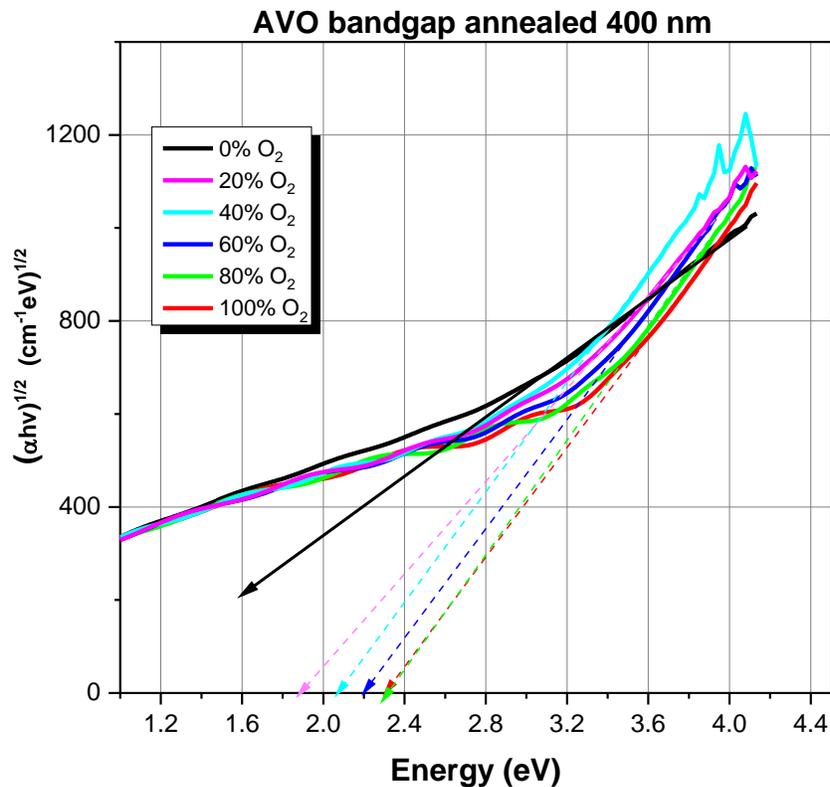

*Figure 5: Tauc plots of various annealed AVO thin films obtained from Transmission and Reflection spectra*

The intrinsic material appears to have an indirect bandgap of approximately in the range of 1.89 – 2.36 eV. Thin films deposited under different reactive plasma exhibits different band gap energies. In metal oxide semiconductors, p orbital from oxygen determines valence band maximum (VBM) and s orbital from metal determines conduction band minimum (CBM) [25]. In pure argon environment deposited thin films, the absorption is poor, and the band gap estimation doesn't corroborate with the approximation. But with 20% oxygen incorporation, it appears that optical bandgap of AVO reduces by uplifting VBM also facilitates hole transfer ability at the interface as seen in photoelectrochemical activities as well.

From the UV-vis spectroscopy, it appears that presence of oxygen in the reactive plasma modulates the indirect band gap of the intrinsic material. In a pure oxygen environment [$O_2/(O_2+Ar) =1$], the estimated band gap of $SbVO_4$ is around 2.36 eV. However, lower oxygen content reduces its band gap monotonically as summarized in the Table 1 below.

| $O_2$/Ar ratio | 0.0 | 0.2 | 0.4 | 0.6 | 0.8 | 1.0 |
|---|---|---|---|---|---|---|
| Band gap (eV) | N/A | 1.89 | 2.14 | 2.2 | 2.36 | 2.36 |

*Table 1: Band-gap variation in RF sputtering deposited $SbVO_4$ with $O_2$ content in the reactive plasma*

Nonetheless, the RF deposited thin films in pure Argon environment displayed low absorption throughout the spectral range. One of the reasons could be because of the presence of unwanted secondary phases in the material deposited without oxygen partial pressure as observed in the X-ray diffraction pattern.

### 5.4 Photoelectrochemical response

The sheet resistance measured using a four-point probe for the annealed RF sputtered film was found to be $7.13 \times 10^8 - 1.1 \times 10^9$ Ω/□, indicating intrinsic insulating nature that arises either because charge transport is significantly restricted or that the carrier densities are very low. An attempt was made to determine the carrier concentrations using capacitance-voltage testing by thermally evaporating 20 nm Au back contacts to AVO films. However, all the devices made by this method were heavily shunted, with resistances on the order of 2-3 Ωcm$^2$. It is unknown whether the shunting is due to the presence of pinholes or the microcracks along the grain boundaries.

The PEC response from representative $SbVO_4$ thin film electrode in three-electrode system is shown in Figure 6 below. This poor PEC performance at lower bias even with high pH electrolyte indicates the presence of hole-trapping sites. It is possible that Sb or V enrichment may lead to induced defects that serve as a recombination center. Literatures suggest that the crystallographic defects such as $V^{5+}$ interstitials and $V^{5+}$ substitution on $Sb^{3+}$ sites can serve as donors [26], a situation contradicted with our lower performing situation with V enrichment. Likewise, defects such as $Sb^{3+}$ or $V^{5+}$ vacancies, or $Sb^{3+}$ substitution on $V^{5+}$ sites can serve as acceptors and result in p-type doping but it still showing no photocurrent supports the argument that non-stoichiometry in the films induce counterproductive defects.

As a consequence, photoexcited holes can only be extracted under high bias. PEC measurements suggest that material to be n-type under higher forward bias against reversible hydrogen electrode (RHE). Issues with AVO include poor hole transport in the bulk, as further evidenced by higher photocurrents for front as opposed to back-side illumination and slow water oxidation kinetics.

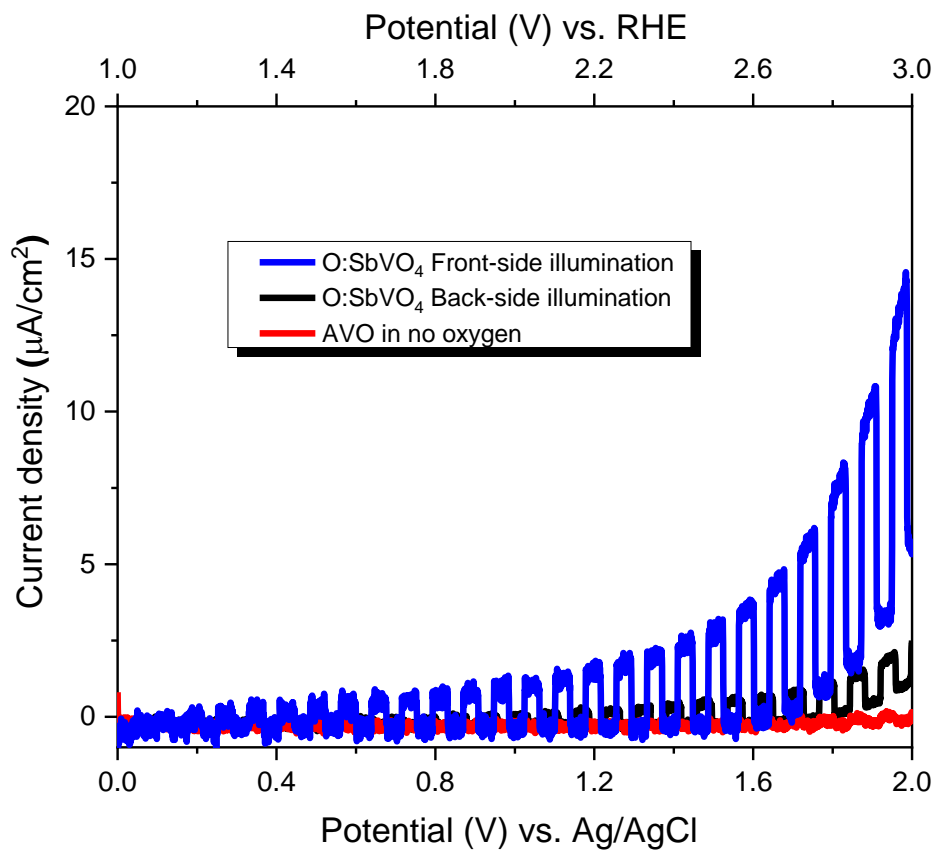

*Figure 6: PEC Performance comparison of O:SbVO$_4$ photoanodes in 0.5M KOH, 5mV/s, 10sec on/off frontside-illumination (blue curve), back-side illumination (black curve), and PEC performance of pure-argon deposited AVO (red curve) in front-illumination.*

The table below shows normalized elemental ratio after running multiple tests under strong alkaline electrolyte of pH 13.2 (0.5 M KOH), and weak acidic electrolyte of pH 4.9 (0.1 M KH$_2$SO$_4$ + 1M Na$_2$SO$_4$). These metal ratios before and after the measurements implies no significant degradation in aqueous media.

Although the RF sputtered AVO exhibits poor photocatalytic activities in lower bias regime, there was not visibly noticeable surface damage or discoloration due to chemical corrosion even after strong bias against the working electrode, demonstrating materials durability and robustness. This may be attributed to the fact that presence of oxygen in plasma produces a more hydrophilic surface via the addition of surface groups [27] [28]. Surface degradation after repeated testing under these conditions were negligible as verified by EDS (as shown in Table 2) after running multiple tests.

| Metal Species | Normalized stoichiometric Ratio (after/before) |
|---|---|
| V | $1.00 \pm 0.02$ |
| Sb | $1.01 \pm 0.01$ |

*Table 2: Normalized Metal Ratios after multiple PEC tests under bias potential in acidic and alkaline media*

**Conclusion:**

Physical vapor deposition (PVD) methods for the synthesis of $SbVO_4$ have never been explored before. The most widely used synthesis methods use sol-gel chemistry coupled with spin-coating or spray pyrolysis to make AVO thin films. Some electrodeposited AVO thin films were also reported.

Sputtering target of $SbVO_4$ can be obtained by sintering and grinding the homogenized mixture of $Sb_2O_3$ and $V_2O_5$ powders. Single phase tetragonal $SbVO_4$ can be prepared under reactive sputtering with controlled oxygen plasma pressure. The surface morphology and band gap modulation of RF sputtered thin film of $SbVO_4$ has been studied. It is found that the presence of oxygen during sputtering efficiently controls the thin film stoichiometry, surface morphology, and modulates the band gap of these oxides as well. The sputtered AVO thin films in reactive plasma environment with 20% oxygen partial pressure $[O_2/(O_2+Ar) = 0.20]$ in 10 mTorr of total pressure showed relatively low band gap of 1.89 eV.

It has been concluded that antimony vanadium oxide is a unique semiconductor material which has potential applications in the conversion of sunlight because of its relatively lower and controllable band gap. Future works to mitigate recombination and improve hole-transfer ability may lead $SbVO_4$ to be used directly in water splitting.

# References

[1]  A. Fujishima and K. Honda, "Electrochemical Photolysis of Water at a Semiconductor Electrode," *Nature,* pp. 37-38, 1972.

[2]  L. Chen, M. E. Graham, G. Li and K. A. Gray, "Fabricating highly active mixed phase TiO2 photocatalysts by reactive DC magnetron sputter deposition," *Thin Solid Films,* pp. 1176-1181, 2006.

[3]  F. Di Franco, M. Santamaria, F. Di Quarto, E. Tsuji and H. Habazaki, "The influence of nitrogen incorporation on the optical properties of anodic Ta2O5," *Electrochimica Acta,* pp. 382-386, 2012.

[4]  G. Wang, H. Wang, Y. Ling, Y. Tang, X. Yang, R. C. Fitzmorris, C. Wang, J. Z. Zhang and Y. Li, "Hydrogen-Treated TiO2 Nanowire Arrays for Photoelectrochemical Water Splitting," *Nano Lett.,* pp. 3026-3033, 2011.

[5]  D. K. Sivula and P. D. M. G. Florian Le Formal, "Solar Water Splitting: Progress Using Hematite (α-Fe2O3) Photoelectrodes," *ChemSusChem,* 2011.

[6]  T. Hisatomi, J. Kubota and K. Domen, "Recent advances in semiconductors for photocatalytic and photoelectrochemical water splitting," *Chem. Soc. Rev.,* pp. 7520-7535, 2014.

[7]  M. G. Walter, E. L. Warren, J. R. McKone, S. W. Boettcher, Q. Mi, E. A. Santori and N. S. Lewis, "Solar Water Splitting Cells," *Chem. Rev. ,* pp. 6446-6473, 2010.

[8]  Z. Zou, J. Ye, K. Sayama and H. Arakawa, "Direct splitting of water under visible light irradiation with an oxide semiconductor photocatalyst," *Nature,* pp. 625-627, 2001.

[9]  V. M. Bondareva, T. V. Andrushkevich and G. A. Zenkovets, "Ammoxidation of Methylpyrazine over Binary Oxide Systems: II. The Study of," *Kinetics and Catalysis,* 1997.

[10] M. M. Bettahar, G. Costentin, L. Savary and J. C. Lavalley, "On the partial oxidation of propane and propylene on mixed metal oxide catalysts," *Applied Catalysis,* 1996.

[11] J. Nilsson, A. L. Canovas, S. Hansen and A. Andersson, "Catalysis and Structure of the SbVO4/Sb2O4 system for propane ammoxidation," *Catalysis Today,* vol. 33, pp. 97-108, 1997.

[12] R. Rahmatollah, M. M. Masoumeh and Z. Solmaz, "SbVO4-TiO2 Cation Deficient Photocatalyst: Synthesis and Photocatalytic Investigation," *Advance Materials Research,* pp. 51-55, 2013.

[13] A. Loiudice, J. Ma, W. S. Drisdell, T. M. Mattox, J. K. Cooper, T. Thao, C. Giannini, J. Yano, L.-W. Wang, I. D. Sharp and R. Buonsanti, "Bandgap Tunability in Sb-alloyed BiVO4 Quaternary Oxides as Visible Light Absorbers for Solar fuel Applications," *Advanced Materials,* pp. 6733-6740, 2015.

[14] G. Centi, P. Mazzoli and S. Perathoner, "Dependence of the catalytic behavior of V—Sb-oxides in propane ammoxidation to acrylonitrile from the method of preparation," *Applied Catalysis A,* 1997.

[15] S. Hansen, K. Stahl, R. Nilsson and A. Andersson, "The Crystal Structure of Sb0.92V0.92O4, Determined by Neutron and Dual Wavelength X-ray Powder Diffraction," *Journal of Solid State Chemistry,* pp. 340-348, 1993.

[16] T. Gron, A. Krajewski, H. Duda and E. Filipek, "The electrical n-p phase transition in the Sb0.92V0.92O4 and Sb2V2O9 compounds," *Journal of Materials science,* vol. 40, no. 19, p. 5299–5301, 2005.

[17] H. zhang, J. Zhang, K. Sun, Z. Feng, P. Ying and C. Li, "Catalytic Performance of the Sb–V Mixed Oxide on Sb–V–O/SiO2 Catalysts in Methane Selective Oxidation to Formaldehyde," *Catalysis Letters,* vol. 106, no. 1-2, pp. 89-93, 2006.

[18] J. Morales, L. Sanchez, F. Martin and F. Berry, "Electrochemical reaction of lithium with nanosized vanadium antimonate," *Journal of Solid State Chemistry,* vol. 179, no. 8, pp. 2554-2561, 2006.

[19] J. F. Brazdil, M. A. Toft, J. P. Bartek, R. G. Teller and R. M. Cyngier, "Sol-Gel Method for Preparing Vanadium-Antimony Oxide Catalysts," *Chem. Mater,* pp. 4100-4103, 1998.

[20] T. Birchall and A. W. Sleight, "Oxidation States in Vanadium antimonate," *Inorganic Chemistry,* vol. 14, pp. 868-870, 1976.

[21] A. R. L. Canovas, J. F. Garcia and S. Hansen, "Structural flexibility in SbVO4," *Catalysis Today,* vol. 158, pp. 156-161, 2010.

[22] G. M. O. Perez and M. A. Banares, "Operando Raman-GC studies of alumina-supported Sb-V-O catalysts and role of the preparation method," *Catalysis Today,* vol. 96, pp. 265-272, 2004.

[23] X. Y. Chen and S. W. Lee, "Controlled synthesis and characterization of colloidal SbVO4 nanocrystals by facile solution-phase method," *Chemical Physics Letters,* vol. 445, pp. 221-226, 2007.

[24] A. F. Fatwa, H. Lihao, A. H. M. Smets, M. Zeman, D. Bernard and R. V. d. Krol, "Efficient solar water splitting by enhanced charge separation in a bismuth vanadate-silicon tandem photoelectrode," *Nature Communicataions,* 2013.

[25] B. Weng, Z. Xiao, W. Meng, C. R. Grice, T. Poudel, X. Deng and Y. Yan, "Bandgap Engineering of Barium Bismuth Niobate Double," *Advanced Energy materials,* 2017.

[26] W.-J. Yin, S.-H. Wei, M. M. Al-Jassim, J. Turner and Y. Yan, "Doping properties of monoclinic BiVO4 studied by first-principles density-functional theory," *Physical Review B,* vol. 83, p. 155102, 2011.

[27] J. Chai, F. Lu, B. Li and D. Y. Kwok, "Wettability Interpretation of Oxygen Plasma Modified Poly(methyl methacrylate)," *Langmuir,* vol. 20, no. 25, pp. 10919-10927, 2004.

[28] F. Walther, P. Davydovskaya, S. Zürcher, M. Kaiser, H. Herberg, A. M. Gigler and R. W. Stark, "Stability of the hydrophilic behavior of oxygen plasma activated SU-8," *Journal of Micromechanics and Microengineering,* vol. 17, pp. 524-531, 2007.

[29] P. Kubelka and F. Munk, "Ein Beitrag zur Optik derFarbanstriche," *Z. Tech. Phys. 12,* p. 593, 1931.

[30] P. Kubelka, "New contributions to the optics of intensely light-scattering materials.," *J. Opt. Soc. Am.,* p. 448, 1948.